\documentclass[%
reprint,
superscriptaddress,
 amsmath,amssymb,
 aps,
 prb,
]{revtex4-2}

\usepackage{graphicx}
\usepackage{dcolumn}
\usepackage{bm}

\usepackage[caption=false]{subfig}
\usepackage{siunitx, soul, xcolor}
\usepackage{adjustbox}
\usepackage{gensymb}
\usepackage[colorlinks=true, allcolors=blue]{hyperref}
\usepackage{orcidlink}

\usepackage{xparse,xcoffins}

\ExplSyntaxOn
\NewCoffin\imagecoffin
\NewCoffin\labelcoffin

\keys_define:nn { miguel/label }
 {
  label   .tl_set:N = \l_miguel_label_tl,
  labelbox .bool_set:N = \l_miguel_label_box_bool,
  labelbox .default:n = true,
  fontsize .tl_set:N = \l_miguel_label_size_tl,
  fontsize .initial:n = \footnotesize,
  pos .choice:,
  pos/nw .code:n = \tl_set:Nn \l_miguel_label_pos_tl { left,up },
  pos/ne .code:n = \tl_set:Nn \l_miguel_label_pos_tl { right,up },
  pos/sw .code:n = \tl_set:Nn \l_miguel_label_pos_tl { left,down },
  pos/se .code:n = \tl_set:Nn \l_miguel_label_pos_tl { right,down },
  pos/n .code:n = \tl_set:Nn \l_miguel_label_pos_tl { hc,up },
  pos/w .code:n = \tl_set:Nn \l_miguel_label_pos_tl { left,vc },
  pos/s .code:n = \tl_set:Nn \l_miguel_label_pos_tl { hc,down },
  pos/e .code:n = \tl_set:Nn \l_miguel_label_pos_tl { right,vc },
  pos .initial:n = nw,
  unknown .code:n   = \clist_put_right:Nx \l_miguel_label_clist
                       { \l_keys_key_tl = \exp_not:n { #1 } }
 }
\clist_new:N \l_miguel_label_clist
\box_new:N \l_miguel_label_box
\box_new:N \l_miguel_label_image_box

\NewDocumentCommand{\xincludegraphics}{O{}m}
 {
  \group_begin:
  \tl_clear:N \l_miguel_label_tl
  \clist_clear:N \l_miguel_label_clist
  \keys_set:nn { miguel/label } { #1 }
  \tl_if_empty:NTF \l_miguel_label_tl
   {
    \miguel_includegraphics:Vn \l_miguel_label_clist { #2 }
   }
   {
    \SetHorizontalCoffin\imagecoffin
     {
      \miguel_includegraphics:Vn \l_miguel_label_clist { #2 }
     }
    \SetHorizontalCoffin\labelcoffin
     {
      \raisebox{\depth}
       {
        \bool_if:NTF \l_miguel_label_box_bool
         { \fcolorbox{white}{white}{\l_miguel_label_size_tl\l_miguel_label_tl} }
         { \l_miguel_label_size_tl\l_miguel_label_tl }
       }
     }
    \SetVerticalPole\imagecoffin{left}{-2pt+\CoffinWidth\labelcoffin/2}
    \SetVerticalPole\imagecoffin{right}{\Width-3pt-\CoffinWidth\labelcoffin/2}
    \SetHorizontalPole\imagecoffin{up}{\Height+5pt-\CoffinHeight\labelcoffin/2}
    \SetHorizontalPole\imagecoffin{down}{3pt+\CoffinHeight\labelcoffin/2}
    \use:x{\JoinCoffins\imagecoffin[\l_miguel_label_pos_tl]\labelcoffin[vc,hc]} 
    \TypesetCoffin\imagecoffin
   }
   \group_end:
 }
\NewDocumentCommand{\setlabel}{m}
 {
  \keys_set:nn { miguel/label } { #1 }
 }

\cs_new_protected:Nn \miguel_includegraphics:nn
 {
  \includegraphics[#1]{#2}
 }
\cs_generate_variant:Nn \miguel_includegraphics:nn { V }

\ExplSyntaxOff

\begin{document}
\preprint{APS/123-QED}

\title{Spin-Charge Conversion in an Epitaxial Kagome Flat-Band Metal}

\author{Wenyi Zhou\,\orcidlink{0000-0001-9717-5955}}
\affiliation{The Ohio State University, Department of Physics, Columbus, OH, USA}
\author{Yuqing Zhu\,\orcidlink{0009-0001-0988-9761}}
\affiliation{The Ohio State University, Department of Physics, Columbus, OH, USA}
\author{Katherine Johnson}
\affiliation{The Ohio State University, Department of Physics, Columbus, OH, USA}
\author{Hari Paudyal\,\orcidlink{0000-0002-3661-0826}}
\affiliation{University of Iowa, Department of Physics and Astronomy, Iowa City, IA, USA}
\author{Duy Nguyễn\,\orcidlink{0009-0001-6060-4098}} 
\affiliation{The Ohio State University, Department of Physics, Columbus, OH, USA}
\author{Dimitri King\,\orcidlink{0009-0004-5359-8314}}
\affiliation{The Ohio State University, Department of Physics, Columbus, OH, USA}
\author{Jaden Ma\,\orcidlink{0009-0008-9045-5151}}
\affiliation{The Ohio State University, Department of Physics, Columbus, OH, USA}
\author{Jay A. Gupta\,\orcidlink{0000-0002-3908-7719}}
\affiliation{The Ohio State University, Department of Physics, Columbus, OH, USA}
\author{Michael E. Flatt\'e\,\orcidlink{0000-0001-5093-1549}}
\affiliation{University of Iowa, Department of Physics and Astronomy, Iowa City, IA, USA}
\author{Roland K. Kawakami\,\orcidlink{0000-0003-0245-9192}}
\email[email address: ]{kawakami.15@osu.edu}
\affiliation{The Ohio State University, Department of Physics, Columbus, OH, USA}

\begin{abstract}
The formation of topological flat bands near the Fermi level in kagome-structured metals has generated substantial research interest. One key consequence of this electronic structure is the strong anisotropy in charge transport, such as the significantly higher in-plane resistivity relative to the out-of-plane direction. However, the spin transport properties of these flat-band kagome metals are still largely unexplored.
In this work, we investigate spin-charge conversion in CoSn, a representative flat-band kagome metal. Scanning tunneling microscopy is employed to image the CoSn surface with atomic resolution, and $dI/dV$ spectroscopy shows clear signatures of the flat bands. Using magneto-optical techniques,
we quantify the spin-orbit torque (SOT) in bilayers of ferromagnetic nickel (Ni) and c-plane CoSn thin films. 
We find a large effective spin Hall angle of $\theta_{SH} = 0.484 \pm 0.005$ within the CoSn layers, significantly exceeding the value measured for Pt of $\theta_{SH} = 0.0436 \pm 0.0003$ in Ni/Pt bilayers.
Density functional theory (DFT) calculations provide a microscopic understanding of the intrinsic spin Hall effect in CoSn, elucidating the sources of Berry curvature in momentum space. These findings establish kagome metals as promising materials for efficient spin-charge conversion and low-power spintronics.
\end{abstract}

\keywords{spin-orbit torque, spin Hall effect, kagome metals, flat bands, magneto-optic Kerr effect}

\maketitle

\section{Introduction} \label{sec:intro}

The two-dimensional (2D) kagome lattice is a subject of intense study due to its inherently fascinating electronic band structure, which features topological Dirac cones and flat bands even within a simple tight-binding framework. When these 2D kagome layers are stacked to form three-dimensional (3D) solids, the resulting band structure retains many of the 2D lattice's key electronic signatures. The presence of these characteristic features has direct experimental consequences. For instance, the Dirac cones in ferromagnetic TbMn$_6$Sn$_6$ give rise to a large anomalous Hall effect observed in experiments, which could evolve into a quantized anomalous Hall effect by tuning the Fermi level into the spin-orbit gap of the Dirac cone~\cite{yin2020quantum}. Similarly, flat bands have been observed in paramagnetic CoSn bulk crystals and epitaxial films through angle-resolved photoemission spectroscopy (ARPES) experiments~\cite{kang2020topological,liu_orbital-selective_2020,huang2022flat,thapaliya2021high,cheng_epitaxial_2023}. Flat bands are interesting because of their anisotropic transport properties and their high density of states, which could lead to unconventional superconductivity and magnetism, as observed in twisted bilayer graphene \cite{cao2018unconventional,serlin_intrinsic_2020}.

\begin{figure}[htbp]
\setlabel{pos=nw,fontsize=\scriptsize} 

\subfloat{\xincludegraphics[width=0.45\textwidth,label=\textcolor{black}{(a)}]{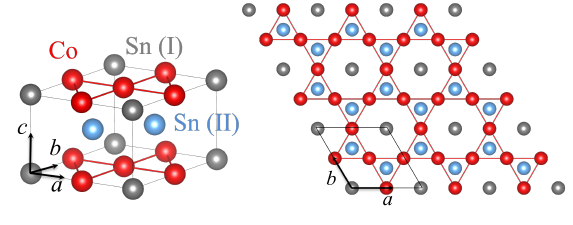}\label{fig:cosn_unit_cell}}
  \hfill
  
  \subfloat{\xincludegraphics[width=0.45\textwidth,label=\textcolor{black}{(b)}]{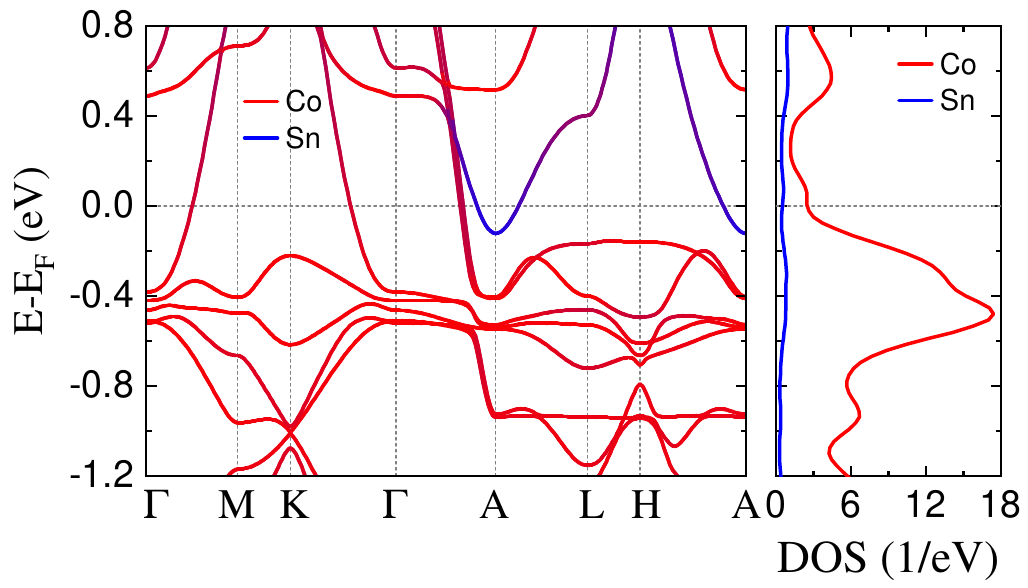}\label{fig:flatband_DFT}}
  \hfill
  \subfloat{\xincludegraphics[width=0.22\textwidth,label=\textcolor{black}{(c)},trim=0pt 0pt 0pt -8pt]{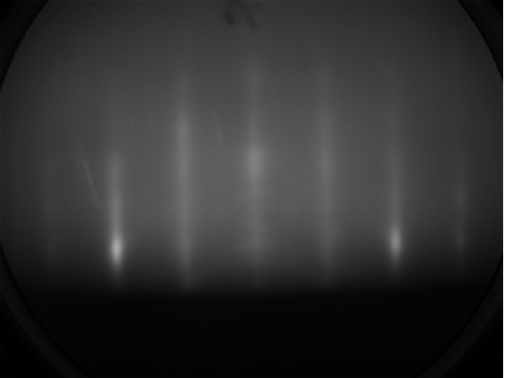}\label{fig:cplane_CoSn_rheed}}
  \hfill
  \subfloat{\xincludegraphics[width=0.22\textwidth,label=\textcolor{black}{(d)},trim=0pt 0pt 0pt -8pt]{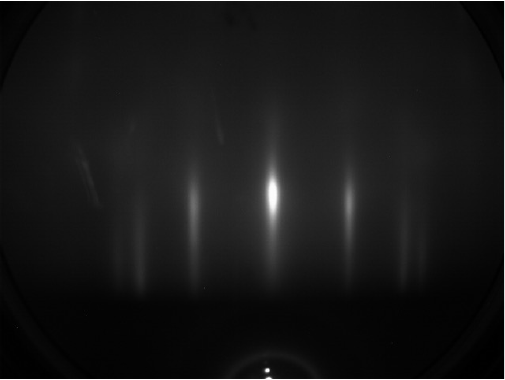}\label{fig:Pt_rheed}}
  \hfill

\caption{\textbf{Sample properties and characterization.} (a) Atomic structure of the CoSn unit cell. (b) Calculated band structure (left) and atom-projected density of states (DOS) (right) for the CoSn, calculated with the inclusion of spin-orbit coupling (SOC). (c) RHEED pattern of a 35 nm $c$-plane CoSn film on an MgO(111) substrate. (d) RHEED pattern of a 20 nm Pt(111) film on an Al$_2$O$_3$(0001) substrate.}\label{fig:cosn_intro}
\end{figure}

Among the kagome materials, paramagnetic CoSn has attracted significant attention. As shown in Figure \ref{fig:cosn_unit_cell}, this material consists of alternating Co$_3$Sn monolayers, where the Co atoms form a kagome lattice (red atoms) and Sn atoms fill the hollow site (grey atoms, labeled ``Sn(I)''), and Sn$_2$ monolayers, where the Sn atoms form a honeycomb lattice (Sn atoms in this layer are labeled ``Sn(II)'' and colored blue). In momentum space, both DFT calculations (Figure \ref{fig:flatband_DFT}) and ARPES measurements experiments~\cite{kang2020topological,liu_orbital-selective_2020,huang2022flat,thapaliya2021high,cheng_epitaxial_2023} identify flat bands near the Fermi level, which could influence the charge transport properties. Indeed, 
the flat bands in CoSn have been shown to cause highly anisotropic charge transport, with the in-plane (flat-band) direction exhibiting a resistivity approximately 60 times larger than the resistivity in the out-of-plane (dispersive) direction~\cite{huang2022flat}.  
However, the spin transport properties of these materials remain largely unexplored. Since CoSn has large spin-orbit coupling and the electronic bands have topological character, it is a candidate for efficient spin-charge conversion and a strong spin Hall effect. 
However, one of the experimental challenges is that epitaxial kagome films are usually grown on metallic buffer layers such as Pt \cite{khadka2020,thapaliya2021high,cheng2021tunable, cheng_atomic_2022}, which itself has a strong spin Hall effect. 
This makes it difficult to separate the contributions from the Pt and kagome layers. We recently succeeded in growing CoSn epitaxial films directly on insulating substrates \cite{cheng_epitaxial_2023}, enabling studies of the spin Hall effect.

In this Letter, we investigate the spin Hall effect in c-plane CoSn epitaxial films through magneto-optic measurements of the spin-orbit torque (SOT) in Ni/CoSn bilayers. 
We find a large effective spin Hall angle of $\theta_{SH} = 0.484 \pm 0.005$ for c-plane CoSn, significantly exceeding the value of $\theta_{SH} = 0.0436 \pm 0.0003$ for Pt(111) determined from SOT measurements in Ni/Pt bilayers.
These studies establish flat-band CoSn as a promising material for efficient spin-charge conversion for potential applications in low-power spintronics.
Calculations based on density functional theory provide insights into the microscopic origin of the spin Hall effect in CoSn, including the sources of Berry curvature in momentum space.

\section{Material Properties,  Sample Preparation, and Characterization} \label{sec:growth}

CoSn crystallizes in the hexagonal $B35$ structure (space group $P6/mmm$, No.~191), as shown in Figure~\ref{fig:cosn_unit_cell}. The crystal structure consists of Cobalt (Co) atoms forming a kagome lattice at the Wyckoff $3f$ sites, with $Sn(I)$ atoms (site $1a$) located at the centers of the hexagons. These kagome layers are stacked along the $c$-axis and interleaved by honeycomb layers of $Sn(II)$ atoms occupying the $2d$ sites. The calculated electronic structure of CoSn, shown in Figure~\ref{fig:flatband_DFT}, captures the characteristic features of a kagome-lattice metal and provides a basis for understanding its spin transport properties. The band dispersion along the high-symmetry path reveals the coexistence of linearly dispersing Dirac-like bands near the $K$ point and a weakly dispersive (nearly flat) band located below the Fermi level ($E_{\rm F}$). The flat band, appearing in the energy range of approximately $-0.2$ to $-0.5$~eV, originates from destructive interference of electronic hopping on the kagome lattice and gives rise to an enhanced DOS, while the Dirac-like crossings in close proximity to $E_{\rm F}$ are expected to generate significant Berry curvature when spin–orbit coupling (SOC) is included. Along the out-of-plane directions, the bands exhibit finite dispersion, indicating a quasi-two-dimensional electronic structure with non-negligible interlayer coupling. The orbital-projected band structure shows that the states near $E_{\rm F}$ are dominated by Co $3d$ orbitals, whereas the Sn contributions are comparatively weak and mainly located away from the Fermi level. The corresponding DOS confirms a finite value at $E_{\rm F}$, consistent with metallic behavior, and exhibits a pronounced peak below $E_{\rm F}$ associated with the flat band. These features, namely the proximity of Dirac crossings and flat bands, are expected to play a central role in generating large intrinsic spin Hall conductivity via enhanced Berry curvature in momentum space, providing a microscopic foundation for the large spin-charge conversion efficiency observed experimentally.

\begin{figure*}
\centering
\includegraphics[width=1\textwidth]{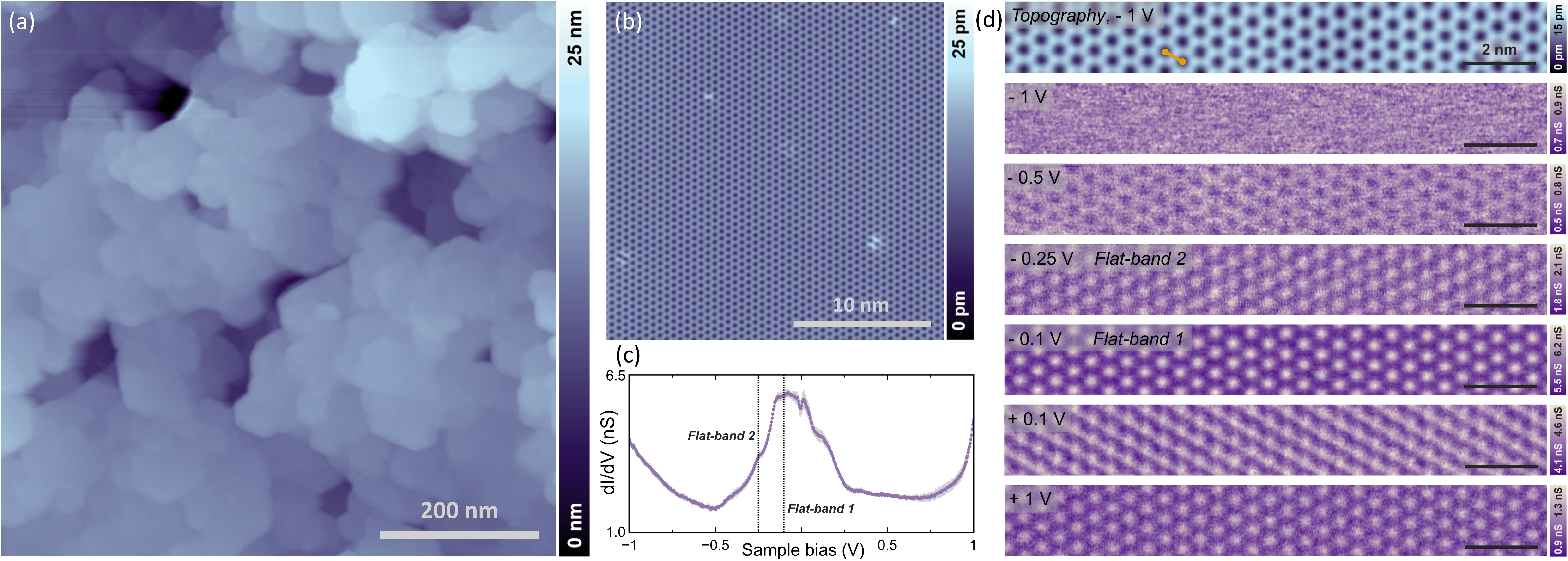}
\caption{
\textbf{STM results of \textit{c}-plane CoSn measured at $4.5$~K.} 
(a) Large-area topograph of a $35$~nm CoSn film on MgO(111) substrate ($V = +1$~V, $I_{\text{set}} = 80$~pA). 
(b) Atomic-resolution image of the Co$_3$Sn-terminated surface ($V = -1$~V, $I_{\text{set}} = 5$~nA). 
(c) Average $dI/dV$ spectrum obtained from ten spectra acquired along the orange line in the top panel of (d). Spectra were taken at fixed height stabilized at $V = -1$~V, $I_{\text{set}} = 6$~nA. Shading denotes the standard deviation across the ten spectra. Dashed lines mark flat-band~1 ($\sim -0.1$~V) and flat-band~2 ($\sim -0.25$~V).
(d) Top: topograph of the mapped region ($V = -1$~V, $I_{\text{set}} = 6$~nA). Below: constant-current $dI/dV$ maps of the same area at the indicated sample biases. 
}
\label{STM}
\end{figure*}

Samples consisting of Ni/CoSn and Ni/Pt bilayers are grown by molecular beam epitaxy (MBE) in an ultrahigh vacuum (UHV) chamber with a base pressure of $4 \times 10^{-10}$ Torr. Thermal effusion cells are used for depositing Co, Sn, Ni, and CaF$_2$. An electron beam evaporator is used for depositing elemental Pt (99.99\% Kurt J. Lesker). 
Samples are characterized \textit{in situ} by reflection high energy electron diffraction (RHEED) (Figures~\ref{fig:cplane_CoSn_rheed} and \ref{fig:Pt_rheed}) to probe the atomic-scale structure of the surface during growth. We cap samples with a transparent 5 nm CaF$_2$ layer at room temperature to prevent oxidation, except for STM samples, which remain uncapped and are transferred in a UHV suitcase.

The growth of c-plane CoSn on MgO(111) follows a three-step process developed in Cheng \textit{et al.}~\cite{cheng_epitaxial_2023}, which is inspired by the Mn$_3$Ge growth on LaAlO$_3$(111) by Hong \textit{et al.}~\cite{hong2022synthesis}. It is worth noting that most of the kagome epitaxial growths are on metal buffer layers \cite{markou2018,taylor2020,hong_large_2020,khadka2020-1,thapaliya2021high}, and the growth of continuous films directly on insulating substrates is relatively unusual \cite{inoue_molecular_2019,hong2022synthesis,cheng_epitaxial_2023,han_molecular_2024}. After a $500 \degree$ C anneal of the MgO(111) substrate in ultrahigh vacuum, a 5 nm CoSn nucleation layer is grown at $470 \degree$C. A 20 nm continuation layer is grown at $100 \degree$C, and a 10 nm finishing layer is grown at $300 \degree$C. Figure \ref{fig:cplane_CoSn_rheed} shows the RHEED pattern of the 35 nm c-plane CoSn film.  The streaky pattern indicates two-dimensional growth with finite terrace width. To measure the resistivity of CoSn, the films are patterned into Hall bars via photolithography and Ar ion milling, and four-probe resistance measurements are performed.
For SOT measurements, a 5 nm Ni overlayer is grown at $50 \degree$C on top of a 35 nm CoSn film. The Ni/CoSn bilayer is patterned into a 20 $\mu$m wide channel that connects two larger contact pads.

The Pt(111) films are deposited on Al$_2$O$_3$(0001) substrates following Cheng \textit{et al.} \cite{cheng_atomic_2022}. Substrates are prepared by annealing in air at $1000 \degree$C for 3 hours followed by annealing in ultrahigh vacuum at $500 \degree$C for 30 min. A 0.6 nm Pt seed layer is grown at $470 \degree$C. The substrate is then lowered to $140 \degree$C, and a 20 nm Pt film is grown while the substrate temperature is gradually lowered to $80 \degree$C. After growth, the film is post-annealed at $400 \degree$C for 20 min.  Figure \ref{fig:Pt_rheed} shows the streaky RHEED pattern of a 20 nm Pt(111) film. The resistivity of Pt is determined through four-probe measurements on patterned Hall bars. For SOT measurements, a 5 nm Ni overlayer is deposited at $50 \degree$C on top of a 20 nm Pt(111) film, and the Ni/Pt bilayer is patterned into a 20 $\mu$m wide channel. 

For STM measurements, a Pt film is deposited on a corner region of the substrate and a small Ta foil clip is placed in contact with the Pt film and spot-welded to the sample paddle prior to growth.
This creates a continuous electrical conduction path between the deposited film and the paddle. The uncapped sample is transferred from the MBE growth chamber to the STM system in a UHV suitcase with a base pressure of ${\sim}10^{-8}$~mbar, so that the surface is never exposed to air. After transfer, the sample is annealed at $150\,^\circ$C for an hour at $\sim 10^{-9}$~mbar to desorb residual adsorbates. STM measurements are conducted at $10^{-11}$~mbar using a CreaTec LT-STM ($4.5$~K) with an electrochemically etched Ni tip. Scanning tunneling spectra ($dI/dV$) are acquired via lock-in detection with an ac bias modulation of $20$~mV at $877$~Hz. Spectra are taken at a fixed tip height, with the feedback loop opened at the start of the sample bias voltage scan. Spatial $dI/dV$ maps are recorded with the feedback loop engaged while the tip is scanned (i.e.~at constant current).

Figure~\ref{STM}a shows a large-area topographic image of the $35$-nm-thick $c$-plane CoSn film. The surface contains coalesced grains with no exposed substrate, consistent with the formation of a continuous film. The grains have lateral dimensions of approximately $20$--$90$~nm, and neighboring grains typically differ in height by less than $10$~nm. Within individual grains, the surface is atomically flat over tens of nanometers. Atomic-resolution imaging (Fig.~\ref{STM}b) reveals a Co$_3$Sn-terminated surface with an in-plane periodicity of $\sim 5.2$~\AA{} and a low density of defects. Figure~\ref{STM}c shows the spatially average $dI/dV$ spectrum acquired along the line marked in the top panel of Figure~\ref{STM}d. The spectrum is metallic and exhibits two occupied-state features: a peak at approximately $-0.1$~V, assigned to flat-band~1, and a weaker shoulder near $-0.25$~V, assigned to flat-band~2. This is in good qualitative agreement with the calculated DOS, here and in prior ARPES and STM/STS studies~\cite{liu_orbital-selective_2020,chen_visualizing_2023}. At these biases, the $dI/dV$ maps in Figure~\ref{STM}d exhibit enhanced conductance at the kagome-lattice centers, which appear as depressions in the simultaneously acquired topographs. This site-selective spectral weight is consistent with the real-space localization pattern previously reported for the kagome flat-band states in CoSn~\cite{chen_visualizing_2023}. The topographs acquired simultaneously at each mapping bias (not shown) exhibit the same atomic-scale contrast pattern as the example shown in the top panel of Figure~\ref{STM}d.

\begin{figure*}
  \subfloat{\xincludegraphics[width=0.32\textwidth,label=\textcolor{black}{(a)}]{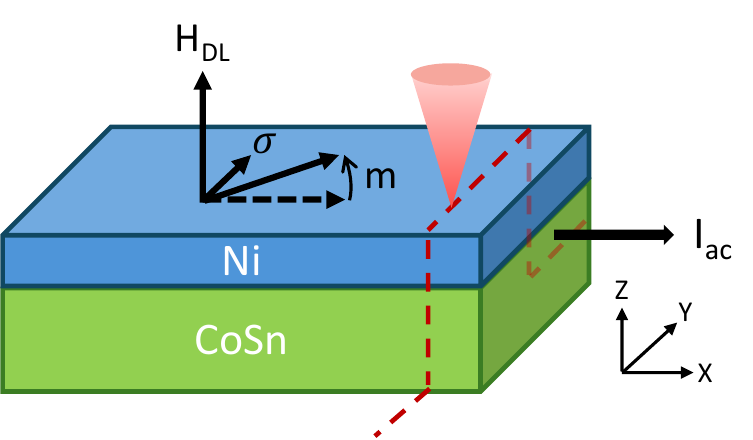}\label{fig:experimental_setup}}
  \hfill
  \subfloat{\xincludegraphics[width=0.32\textwidth,label=\textcolor{black}{(b)}]{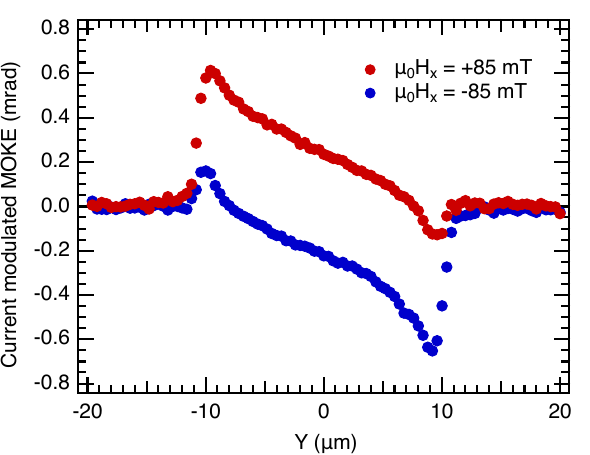}\label{fig:MOKE_raw}}
  \hfill 
  \subfloat{\xincludegraphics[width=0.32\textwidth,label=\textcolor{black}{(c)}]{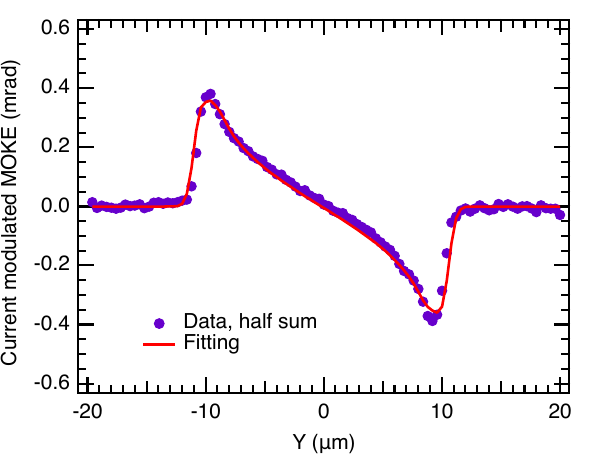}\label{fig:MOKE_Oersted}}
  \hfill
  \subfloat{\xincludegraphics[width=0.32\textwidth,label=\textcolor{black}{(d)}]{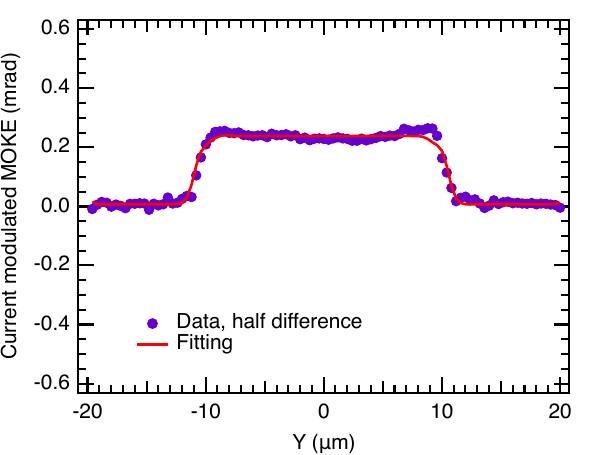}\label{fig:MOKE_DLSOT}}
  \hfill
  \subfloat{\xincludegraphics[width=0.32\textwidth,label=\textcolor{black}{(e)}]{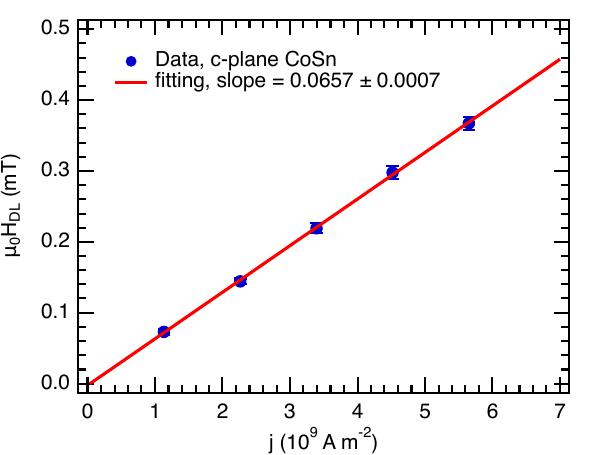}\label{fig:MOKE_HDLvsj_CoSn}}
  \hfill
  \subfloat{\xincludegraphics[width=0.32\textwidth,label=\textcolor{black}{(f)}]{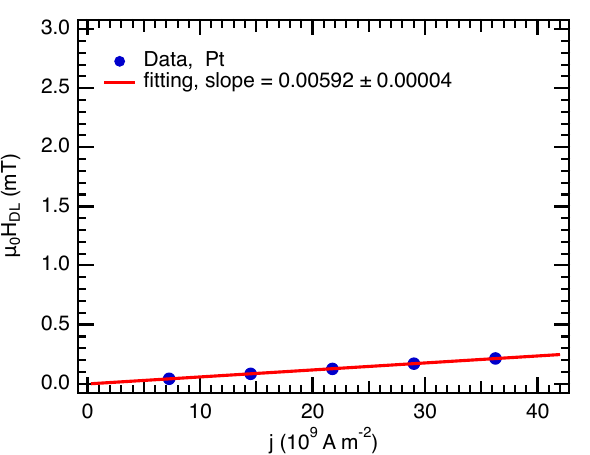}\label{fig:MOKE_HDLvsj_Pt}}
  \hfill

   \caption{\textbf{DL-SOT detected by current-modulated polar MOKE.}
  (a) Schematic of experimental setup for using polar MOKE to detect the DL-SOT. An ac current through the sample along the $x$ direction generates DL-SOT, which rotates the magnetization a small angle out-of-plane, which is detected by MOKE using a lock-in amplifier.
  (b) An example of a line scan across the channel of Ni/CoSn (5/35\,nm) device, with the ac current density $j_c = 5.65 \times 10^{9}$ A/m$^{2}$ applied to the channel.
  (c) The half sum $[\theta_K(+85$ mT$) + \theta_K(-85$ mT$)]/2$ of the current modulated MOKE voltages shown in (b) that corresponds to the out-of-plane Oersted field.
  (d) The half difference of the current modulated MOKE voltages shown in (b) $[\theta_K(+85$ mT$) - \theta_K(-85$ mT$)]/2$ measures the effective DL-SOT field, $H_{DL}$. 
  (e) The linear dependence of effective DL-SOT field vs.~ac current density in Ni/CoSn, yields a slope of $\mu_0 H_{DL}/j = (65.7 \pm 0.7)$ mT per $10^{12}$ A/m$^{2}$. (f) The linear dependence of effective DL-SOT field vs.~ac current density in Ni/Pt, yields a slope of $\mu_0 H_{DL}/j = (5.92 \pm 0.04)$ mT per $10^{12}$ A/m$^{2}$. For comparing the slopes in panels (e) and (f), the ranges for the axes maintain the same aspect ratio.
}\label{fig:MOKE_figure}
\end{figure*}

\section{Measuring the Spin-Charge Conversion Efficiency} \label{sec:properties}

Figure~\ref{fig:experimental_setup} shows the magneto-optical measurement of the spin-charge conversion efficiency, following Fan \textit{et al.}~\cite{fan2014} and Lyalin \textit{et al.}~\cite{lyalin_spinorbit_2021}, and is outlined in the following. In a bilayer of a ferromagnetic (FM) layer (e.g., Ni) and a spin-orbit (SO) layer (e.g., CoSn, Pt), a current flowing in the SO layer produces an out-of-plane spin current via the spin Hall effect (or Rashba-Edelstein effect). 
The spin polarization $\boldsymbol{\sigma}$ is typically perpendicular to the charge current, and it produces a damping-like spin-orbit torque (DL-SOT) that causes the in-plane magnetization of the FM to rotate slightly out-of-plane, generating a $z$ component of magnetization. This can be viewed as an effective magnetic field $\mathbf{H_{DL}} \sim \mathbf{m} \times \boldsymbol{\sigma}$ oriented out-of-plane, where $\mathbf{m}$ is the unit vector of magnetization. We measure $m_z$ using the magneto-optic Kerr effect (MOKE) in polar geometry (normal incidence), where the linearly-polarized incident beam experiences a rotation of its polarization axis (Kerr rotation $\theta_K$) that is proportional to $m_z$ upon reflection from the surface. As discussed later,  the value of $\theta_{K}$ is converted to the strength of the DL-SOT ($\mathbf{H_{DL}}$), and this is directly related to the spin-charge conversion efficiency in the SO layer, typically expressed as the effective spin Hall angle $\theta_{SH} \equiv {j_s}/{j_c}$ (ratio of spin and charge current densities).

Figures~\ref{fig:MOKE_raw} - \ref{fig:MOKE_HDLvsj_Pt} show representative data and analysis for a Ni/c-plane CoSn (5/35\,nm) bilayer sample. The film is patterned to a 20\,$\mu$m wide channel, and the laser is scanned across the channel to obtain a linecut of the Kerr rotation ($\theta_K$ vs.~$y$ position).
A mode-locked Ti:sapphire laser ($\lambda$ = 800\,nm) is used for polar MOKE measurement to detect $m_z$. The beam is focused to a spot size of $\sim$2\,$\mu$m with an incident power of 600\,$\mu$W. Kerr rotation is detected via polarization analysis using a half-wave plate and Wollaston prism, followed by a balanced photodetector. The signal is then processed by a lock-in amplifier that demodulates at the first harmonic of the sinusoidal current applied at 41227 Hz.
Figure~\ref{fig:MOKE_raw} shows the current-modulated MOKE linecut at positive saturation of the ferromagnet $\theta_K(+H_x)$ (red points) and at negative saturation $\theta_K(-H_x)$ (blue points) for an ac current density of $j_c = 5.65 \times 10^{9}$ A/m$^2$ (rms) in the CoSn layer. 

The signals originate from a combination of the DL-SOT effective field $H_{DL}$ and the out-of-plane Oersted field $H_{Oe,z}$ originating from Ampere's Law. These contributions can be separated because they have different symmetries with respect to the magnetization, where $\mathbf{H_{DL}}\sim\mathbf{m}\times\boldsymbol{\sigma}$ flips with $\mathbf{m}$, while $\mathbf{H_{Oe}}$ is independent of $\mathbf{m}$. 
Thus, the ``half sum,'' $[\theta_K(+H_x)+\theta_K(-H_x)]/2$, shown in Figure \ref{fig:MOKE_Oersted} comes from $H_{Oe,z}$ and the ``half difference,'' $[\theta_K(+H_x)-\theta_K(-H_x)]/2$, shown in Figure \ref{fig:MOKE_DLSOT} comes from $H_{DL}$. In Figure \ref{fig:MOKE_Oersted}, using Ampere's Law to calculate the $z$ component of the Oersted field across the channel and scaling it (red curve) to match the experimental data for the half sum linecut (purple dots in Figure \ref{fig:MOKE_Oersted}) yields a conversion factor of 1.55 Tesla/rad. Applying this conversion factor to the half difference data in Figure~\ref{fig:MOKE_DLSOT} yields a quantitative value of $\mu_0 H_{DL} = 0.37 \pm 0.01$ mT, extracted by fitting with a theoretical curve that is uniform across the channel and zero outside. Both the fitting curves in Figures \ref{fig:MOKE_Oersted} and \ref{fig:MOKE_DLSOT} have been convolved with a Gaussian of width 2\,$\mu$m to account for the size of the focused laser spot. Repeating this procedure for different values of $j_c$ yields a linear dependence of $\mu_0 H_{DL}$ on $j_c$, as shown in Figure \ref{fig:MOKE_HDLvsj_CoSn}. A linear fit of the slope yields a DL-SOT effective field per current density of $65.7 \pm 0.7$ mT per $10^{12}$ A/m$^{2}$.

Assuming that the SOT is generated by the spin Hall effect, we determine the effective spin Hall angle $\theta_{SH}$ using the equation~\cite{ando2008,liu2011}:
\begin{equation}
\dfrac{\mu_0 H_{DL}}{j_c} = \dfrac{\hbar}{2 e} \dfrac{\theta_{SH}}{ M_s t_{FM}},
 \label{eq:H_eff_DL-SOT}
\end{equation}
where $M_s$ is the saturation magnetization of Ni ($M_s = 485$ kA/m~\cite{hou2017,ye2018,khadka2020}) and $t_{FM}$ is the thickness of the Ni layer (5 nm).
The effective spin Hall angle in the Ni/c-plane CoSn bilayer is found to be $\theta_{SH} = 0.484 \pm 0.005$, where the error represents the uncertainty of the fitting. We note that the uncertainty of the layer thickness $t_{FM}$ in equation \ref{eq:H_eff_DL-SOT} is typically a few percent, which produces a larger overall uncertainty in $\theta_{SH}$. 
We call $\theta_{SH}$ the effective spin Hall angle because it is determined not only by the spin Hall angle, but also by the spin transparency of the Ni/c-plane CoSn interface and a damping-like SOT generated at the interface (e.g., a possible damping-like contribution from the Rashba-Edelstein effect).

For comparison, we quantify the spin-charge conversion efficiency in Ni/Pt following the same procedure.
Figure \ref{fig:MOKE_HDLvsj_Pt} shows the damping-like torque as a function of current density for Ni/Pt(111) (5/20\,nm). This yields an effective spin Hall angle of $\theta_{SH} = 0.0436 \pm 0.0003$. This value is consistent with other studies of the effective spin Hall angle in Pt \cite{liu2011,wang2014determination,manchon2019,lyalin_spinorbit_2021}. Notably, the effective spin Hall angle in Ni/CoSn is roughly eleven times larger than for Ni/Pt. This result establishes CoSn as a promising material for spintronics and charge-spin interconversion.

\section{Theoretical Analysis and Discussion} \label{sec:discussion}
To examine the intrinsic contribution to spin--charge conversion in CoSn, we performed density functional theory (DFT) calculations using the Vienna \textit{Ab initio} Simulation Package~\cite{furthmuller1996dimer, kresse1996efficiency} including spin--orbit coupling (SOC), followed by Wannier interpolation using the Wannier90~\cite{Pizzi2019s} and Wannier tool~\cite{WannierTools2017} packages. The calculations were performed using projector augmented wave
pseudopotentials~\cite{blochl1994projector} and Perdew-Burke-Ernzerhof (PBE) functionals~\cite{perdew1996o} within the generalized gradient approximation. The resulting Wannier Hamiltonian was used to evaluate the spin Berry curvature and intrinsic SHC within the Kubo formalism using WannierTools~\cite{guo2008intrinsic, WannierTools2017}.

\begin{figure}[h]
\setlabel{pos=nw,fontsize=\scriptsize} 
  \subfloat{\xincludegraphics[width=0.40\textwidth,label=(a)]{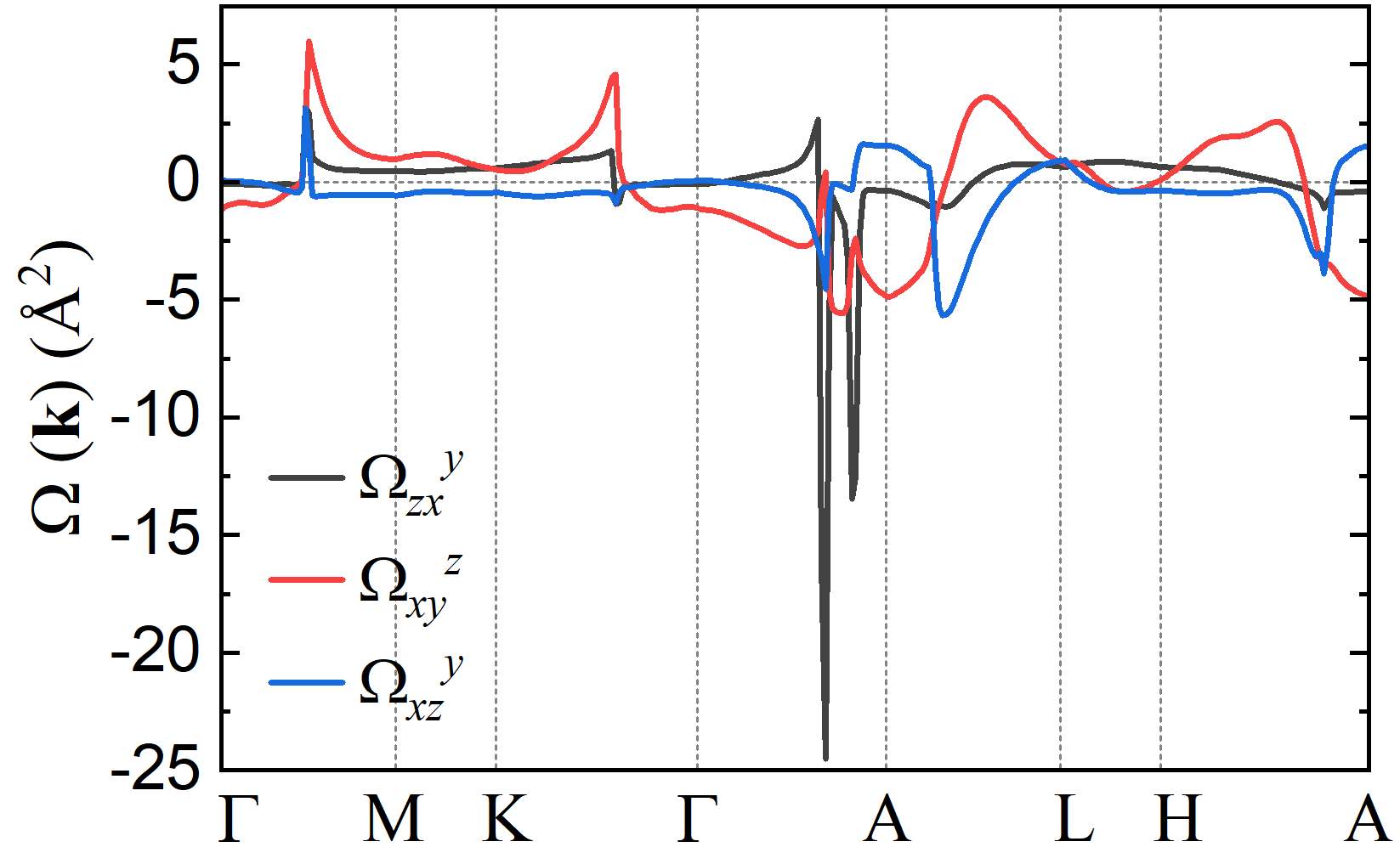}\label{fig:theory_a}}
  \hfill
  \subfloat{\xincludegraphics[width=0.40\textwidth,label=(b)]{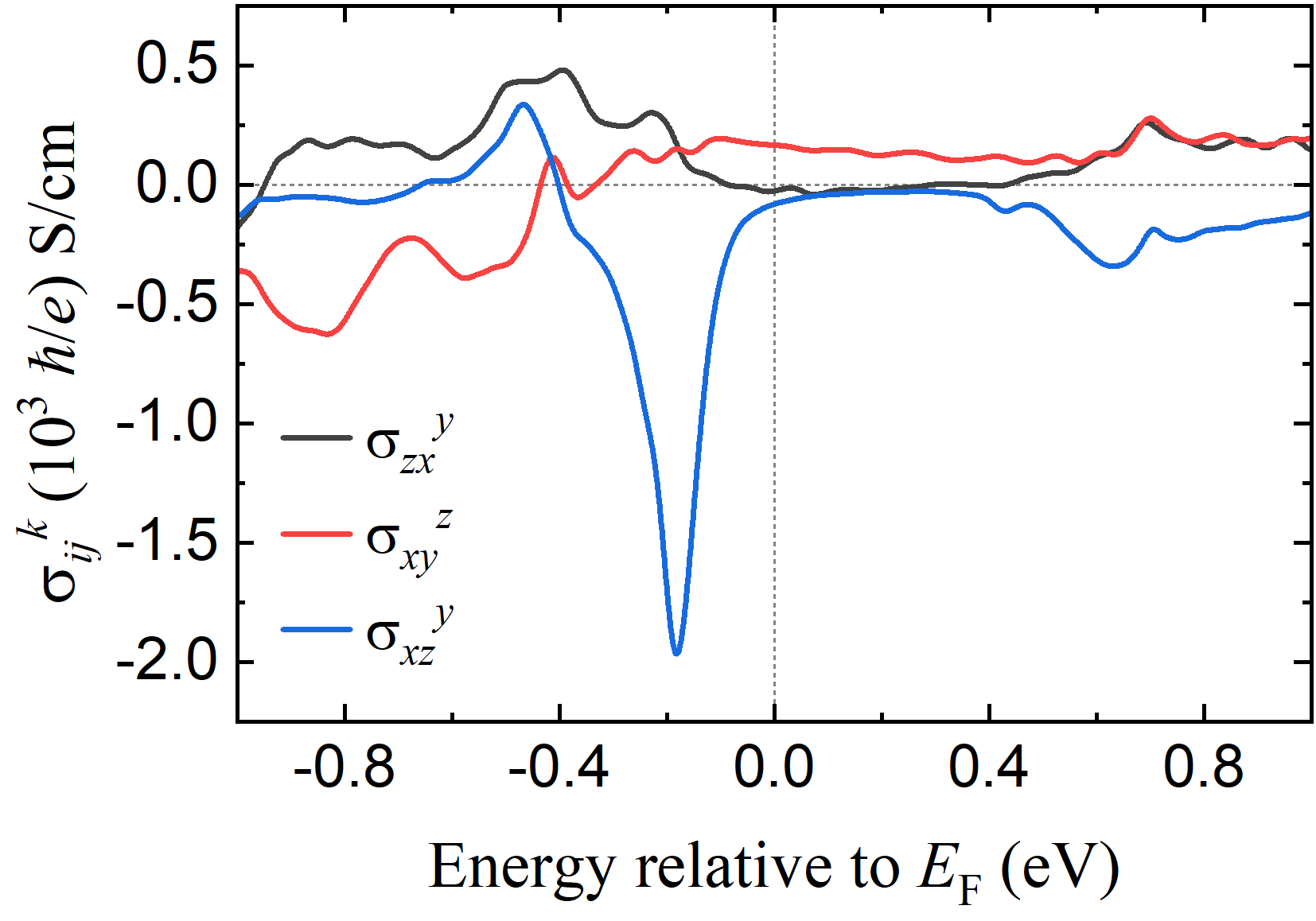}\label{fig:theory_b}
  }
  \hfill
\caption{Spin Berry curvature and intrinsic spin Hall conductivity (SHC) of CoSn calculated from the Wannier-interpolated electronic structure including spin--orbit coupling (SOC). (a) Occupied-state spin Berry curvature components
evaluated at the Fermi energy. The pronounced momentum dependence, including sharp features along the $\Gamma$--A direction, reflects SOC-mediated interband mixing in the CoSn electronic structure. (b) Corresponding intrinsic SHC tensor components 
as a function of energy relative to the Fermi level. 
The strong energy and tensor dependence of the calculated SHC demonstrates the sensitivity of the intrinsic spin Hall response to the electronic structure near the Fermi level.}
\end{figure}

Figure~\ref{fig:theory_a} shows the occupied-state spin Berry curvature components $\Omega_{zx}^{y}$, $\Omega_{xy}^{z}$, and $\Omega_{xz}^{y}$ along the high-symmetry path. All three components exhibit pronounced momentum dependence, with sharp features consistent with SOC-mediated interband mixing in the kagome-derived electronic structure of CoSn. Particularly prominent structure is found in $\Omega_{zx}^{y}$ along the $\Gamma$--A direction. However, the intrinsic SHC depends on contributions from occupied states throughout the Brillouin zone, where different bands and momentum regions may contribute with opposite signs. Consequently, pronounced local spin Berry curvature does not necessarily produce a large integrated spin Hall response.

The corresponding intrinsic SHC components are shown in Figure~\ref{fig:theory_b} as a function of chemical potential relative to the calculated Fermi level. The results reveal substantial tensor anisotropy and energy dependence. In particular, $\sigma_{xz}^{y}$ exhibits a large negative peak below $E_{\rm F}$, whereas $\sigma_{zx}^{y}$ develops a broad positive enhancement at lower energies. The latter is directly relevant to the experimental geometry, where an electric field along $x$ generates a spin current propagating along $z$ with spin polarization along $y$. Thus, the calculated $\sigma_{zx}^{y}$ establishes an intrinsic spin Hall channel capable of generating the damping-like torque observed experimentally. Although this component is relatively modest at the calculated Fermi level, its strong dependence on chemical potential suggests that the intrinsic response may be sensitive to carrier concentration and band filling.

The calculations therefore establish an intrinsic, Berry-curvature-mediated contribution to the spin Hall response of CoSn and reveal substantial anisotropy among the different tensor components. However, the calculated bulk intrinsic SHC should not be identified directly with the experimentally determined effective spin Hall angle of the Ni/CoSn bilayer. The latter can additionally contain contributions from interface spin transmission, interfacial spin--orbit coupling, and extrinsic scattering mechanisms. The combination of these effects with the intrinsic response may contribute to the large spin--charge conversion efficiency observed experimentally.

\section{Conclusion} \label{sec:conclusion}

In summary, we have demonstrated highly efficient spin-charge conversion in c-plane CoSn thin films grown directly on insulating MgO(111) substrates. By employing current-modulated polar magneto-optical Kerr effect measurements on Ni/CoSn bilayers, we determined a large effective spin Hall angle of $\theta_{\text{SH}} = 0.484 \pm 0.005$, approximately eleven times greater than that measured in control Ni/Pt devices ($\theta_{\text{SH}} = 0.0436 \pm 0.0003$). DFT calculations attribute this enhanced spin Hall effect to concentrated Berry curvature hotspots generated by spin-orbit coupling near topological Dirac crossings and flat bands in close proximity to the Fermi level. Overall, these results highlight flat-band kagome metals as a promising material platform for low-power spintronic devices and efficient spin-orbit torque.

\section*{Acknowledgements}
\label{sec:acknowledgements}

This work was supported by the US Department of Energy Grant No.~DE-SC0016379. W.Z., Y.Z., and R.K.K. performed the magneto-optical and transport measurements and analyzed the data. K.J., D.K., J.M., and R.K.K. performed the MBE growth. H.P. and M.E.F. calculated the electronic band structure and spin Hall conductivity. D.N. and J.A.G. performed the STM measurements and analyzed the data. All authors contributed to the manuscript.

\bibliography{CoSn.bib}
\end{document}